\documentclass[fleqn]{article}

\newcommand{\titel}
{Bitcoin: a Money-like Informational Commodity}

\usepackage{stmaryrd,amssymb,amsmath,amsthm,url}

\theoremstyle{definition}

\title{\titel}

\author{
	Jan A.\ Bergstra \& Peter Weijland
	 \\
\\
  {
	  Informatics Institute,
	  University of Amsterdam}\\
	{ Email: \url{j.a.bergstra@uva.nl}, \url{w.p.weijland@uva.nl}
	}
}

\begin{document}
\maketitle

\begin{abstract}
The question ``what is Bitcoin" allows for many answers depending on the objectives aimed at when 
providing such answers. The question addressed in this paper is to determine a top-level classification,
or type, for Bitcoin. We will  classify Bitcoin as a system of type  money-like informational commodity (MLIC). 
\\[5mm]
\end{abstract}

\newpage
{\small\tableofcontents}~\newline~\newline

\section{Introduction}
For an artifact $X$ we will denote with BT($X$) the class of base types of $X$. 
We do not claim that each artifact $X$ has a base type. 
A base type of $X$ should express a characteristic property or objective for $X$.
We intend to apply our analysis in the case that $X$= Bitcoin, and we
 will assume that Bitcoin is an artifact.\footnote{%
 Obviously, the artifact called Bitcoin is a carrier for an activity (process or system) that involves human 
 agents  as well, and which for that reason cannot be understood as an artifact.}

We will use a notational distinction between $X$, a term from some syntax, and $m(X)$, the meaning of $X$ 
in the world. Most plausibly $m(X)$ is not a term from some syntax. 
The question ``what is $m($Bitcoin$)$" is not easy to answer. 
Informatics provides no answers to such questions because it is initially unclear which aspects of 
Bitcoin, when viewed as an operational distributed software system with an evolving 
human user community, must be taken into account.

Base types for $X$ need not be unique. For instance if $X$ is an airplane (say it is of type $A$ for airplane) 
and at the 
same time $X$ is a propellor airplane (say it is of type type $A_p$) then both $A$ and $A_p$ 
may be considered

to be base types for $X$ (that is members of  BT($X$)). In this case $A_p$ is a subtype of $A$. 
Evidently different base types for $X$ must have a non-empty intersection because  $m(X)$
is included in the extension of each base type for $X$. Extensions of base types of $X$ are partially ordered,
but not necessarily totally ordered, by the subset relation for their extensions.

An optimal base type, if it exists, is concise, that is its description is simple, and at the same time it is informative.
These two criteria work in opposite directions. Finding an optimal base type for an artifact class\footnote{%
Instead of artifact class one may prefer to speak of an artifact kind. For a discussion of monies from the 
perspective of natural kinds and artifact kinds  see~\cite{BergstraL2013}.}
may be impossible or unrewarding. For instance looking for a base type for ``organization" is probably pointless,
because ``organization" is best seen as a base type itself.

We will be interested in the case where $X$= Bitcoin.

\subsection{Looking for an ontology of monies}
Our contribution aims at a conceptualization of a part of the area of money. A conceptualization may precede 
formalization in an ontology (for instance making use of the ologs of~\cite{SpivakK2011}), 
and codification of the ontology in dedicated notation such as OWL (\cite{HorrocksPH2003}). Conventional money, 
informational money, and money-like commodities, together span a wealth of possibilities for which a sound
and stable ontological framework can be developed and combined with existing ontological frameworks such as 
the enterprise ontology of~\cite{UscholdKMZ1998}.
\subsection{Defining versus typing}
When considering a class term $Y$ for some kind of entities with $Y$ still vaguely specified by means 
of keywords or combinations thereof, the objective may arise to define ``being a $Y$" in a more rigorous fashion. 
We write $|Y|$ for the extension of $Y$, that is the class of all of its members. In practice membership of 
$|Y|$ may be a matter of degree, plausibility, or probability. For instance with $Y$= ``car" (a word serving as a class term), $|Y|$ is the collection of all cars.\footnote{%
We can talk about cars without having made up our mind about the ``carness" of a wreck. The latter
issue being deferred to the phase of precisely defining cars, preferably guided by having an objective for giving
that definition at hand.}
The extension $|Y|$ of $Y$ may vary in time, 
we write $|Y|^{t}$ for the extension of $Y$ at time $t$.

Now one may wish to consider a particular definition $Y_d$ of $Y$. We write 
$| Y_d |$ for the extension of 
$Y_d$. Conceivably $|Y_d|$  turns out not to have the same
extension as $Y$ (that is to differ from $|Y|$) because striving for precision and 
conciseness entails simplifications that go hand in hand with minor modifications
of the initially assumed extension of $Y$.

\subsubsection{Comparing candidate definitions}
Arguably a phase may exist in which different definitions, say $Y_{d_1}$ and $Y_{d_2}$, of $Y$ are 
viewed as candidates for being given the status of a preferred definition. 
Candidate definitions may be compared
from different angles: conciseness of description, intelligibility, 
similarity with definitions for related concepts, compliance
with standards, conventions and formats, and the proximity of $|Y_{d_i}|$ with the intended extension $|Y|$.

\subsubsection{Defining a single entity}
Clearly $Y_d$ provides a criterion that precisely the members of $|Y_d|$ satisfy. 
If we intend to define a single entity $X$ then
we proceed to define the singleton class $Y_X$ such  that $|Y_X|$ contains only $m(X)$. 
Not all entities $e$ have an informative definition. 
Consider
the natural number $e=257$. In principle we may define $e$ as the unique member of the class defined by
``a natural number reachable after 258 steps, counting from below, and starting with zero". This definition is
hardly informative, however,  and it may be considered circular in an unfortunate way.

When considering Bitcoin we find that we may be looking at a singleton class. 
Assuming that Bitcoin is the process of use and maintenance of a particular evolution of an 
implementation of ``a given specification of Bitcoin" we find that Bitcoin 
may not have a definition in the same way as 257 fails to have one.

In~\cite{Bergstra2013a} one may find some meta-theory of definitions, providing a hierarchy of forms of definitions
meant to be of use for defining the concept of money. We feel that a definition of Bitcoin, 
if it can be given at all, would, just as for the concept of money, requires as a basis the availability of
some explicit meta-theory of definition. That meta-theory won't provide a 
definition of the concept of a definition, however, if uninformative circularity is to be avoided. 
Unavoidably those who agree
on concepts defined by means of ``definitions" need to agree on ``what is a definition" laid down in less rigorous ways.

\subsubsection{Typing}
Typing of an entity $X$ as an alternative to defining an entity $X$ proceeds by determining a so-called type $T$ for $X$. 
Now the extension $|T|$ of $T$ should constitute a class of sufficiently $X$-like entities so that the assertion that 
$X$ is a $T$
makes much sense as an initial piece of information about $X$. Different types of the same $X$ may provide 
further information about it. With a base type of $X$ we will denote a type for $X$ which can be used as a first characterization of ``what kind of entity $X$ is".

Providing a type for $X$ may require the introduction of a new type description. 
That corresponds to the result of finding
a definition of a class with a larger extension than the entity $X$ alone.

\subsubsection{Describing}
We will use description as a term to denote any manner for indicating a class of entities of whatsoever kind. Typically one would start out 
with a description of $X$ and then proceed towards either proposing a type for $X$ or a definition for $X$. Obviously
in this approach the meta-theory of definition won't apply to descriptions.

\subsection{Problem statement}
Our question is (i) to argue that BT(Bitcoin) is non-empty, and (ii) to determine an optimal base type 
(or preferred base type), named OBT(Bitcoin), in BT(Bitcoin). A type B in BT(Bitcoin)
can be used to explain what kind of thing Bitcoin is in an initial explanation from first principles of it. 
We do not claim
that on a priori grounds the existence of a satisfactory answer to this 
question is guaranteed.\footnote{%
The original paper on Bitcoin is~\cite{Nakamoto2008}.
We refer to the paragraph ``Taking Bitcoin Seriously" of~\cite{BergstraL2013b} for a statement on the 
risks of considering Bitcoin from a scholarly perspective. For a 
technical survey of Bitcoin  see e.g.~\cite{Entrup2013}. For some legal information on Bitcoin we refer 
to~\cite{Grinberg2012}. We are not convinced that trying to classify 
Bitcoin ``as a money" requires reconsidering the very concept of money (a viewpoint 
found in~ \cite{MaurerNS2013}). An exploration of the concept of money may be found in~\cite{Bergstra2013a}, 
where special emphasis is given to the notion of a money of account of which a virtual money is considered 
a special case.}

\subsubsection{Methodological difficulty}
The problem of finding an optimal type OBT($X$) in BT($X$), for some kind $X$, 
can only be stated if some description of X
has already been found. This suggests the presence of an unfortunate circularity in our problem statement. In fact
there is no circularity because we may assume that $X$ is a candidate member of BT($X$), it might even be
evaluated as being optimal.

Nevertheless we must assume that $X$ has been characterized in a preliminary 
fashion by means of one or more descriptions that characterize to some extent ``what $X$ is", 
that is a preliminary indication of $|X|$. The difficulty is that
the question seems to be stated in terms of its answer. Rather than indicating a circularity, it appears that
the mentioned difficulty indicates that  the problem of finding an optimal type for an entity is an optimization problem,
that may admit a stepwise solution starting with an initial candidate solution.

\subsubsection{Initial kind descriptions}
Let IKD($X$), the initial kind description of $X$, be a reasonably clear indication of
which $Y$'s are in of the same kind as $X$. IKD($X$) may be quite verbose, 
it may be too technical still requiring a useful abstraction,
it may make use of analogies, it may be a heterogeneous collection of different 
characterizations al of which are supposed to pertain somehow to those $Y$'s 
which are supposed to be of the same kind as $X$. At the same time we allow 
for the situation that the description is redundant as well as marginally inconsistent. 
We call an IKD initial because that kind of description
is a starting point of the search for an (optimal) element in IKD($X$).

In the case of Bitcoin candidates of an IKD abound in the literature. 
Below we propose an IKD for Bitcoin. 
Our proposal for IKD(Bitcoin) represents a self-made combination of statements about 
Bitcoin regularly found on the internet.

\subsubsection{IKD(Bitcoin), an initial  kind description of Bitcoin}
IKD(Bitcoin) reads thus:
Bitcoin is a remarkably successful  P2P system, mainly consisting of open source clients, 
which exists since early 2009, 
having been started up by (at the time of writing) an anonymous programmer 
or group of programmers. As a tool
Bitcoin provides its clients a cryptology based informational money. 
Amounts in the system are expressed in the unit BTC. 
The essential feature of Bitcoin is that consists of a P2P network only and at the same time 
prevents double spending effectively.
Bitcoin is also claimed to allow its users a high degree of anonymity. 
That virtue for the system has been contested by various authors since 2012.

\subsubsection{Moneyness and money-likeness}
Moneyness is the circumstance that a certain system represents a money (e.g. see~\cite{Laumas1968}). 
Rather than a binary variable taking values yes and no, moneyness is a matter of degree. 
With high or significant moneyness we mean that an artifact class could, in principle, be used as a money. 
The artifact is money-like if its functionality resembles that of a money, though only in part.

We assume that a money-like system or artifact may be further removed from a money than an 
artifact with a moderate degree of moneyness. Thus we will assume ``money-likeness" may be 
present in  some artifact class to some significant degree while moneyness is not.

Claiming that a system or artifact $X$ is money-like requires taking some position concerning the 
functions of money which $X$ realizes. It is plausible that the money of exchange function is
considered indispensable while the money of account function is considered less prominent.  Ownership 
(or possession) of quantities of $X$ must bring with it a bundle of rights resembling that of 
a quantity of (well-recognized) money. For instance selling one's holding of $X$ at any time to anyone
must be a right.

\section{Criticizing proposed types for Bitcoin}
We will first list some proposals for typing Bitcoin and we will 
argue why these proposals should be rejected.
\subsection{IKD(Bitcoin)}
IKD(Bitcoin)  does not qualify as an element of BT(Bitcoin) for at least  the following reasons:
\begin{enumerate}
\item Mentioning double spending attacks should not enter OBT(Bitcoin) though it might be mentioned in some types in BT(Bitcoin). We don't claim that double spending might be permitted, but we merely claim 
that singling out one of many possible
failures is premature when designing an OBT fair any artifact (including Bitcoin).
\item Success is not implied by any optimal base type. Successful artifacts in OBT(Bitcoin) constitute a proper 
subtype OBT$_s$(Bitcoin) of OBT(Bitcoin).
\item The year of birth of Bitcoin is an unnecessary detail, which is not an expected entry in any type
description in BT(Bitcoin).
\item It being cryptology based is reasonably considered an unnecessary refinement.
\item No mention is made of the status problem: is Bitcoin a money. 
The suggestion that Bitcoin's status is ``money" follows too easily from the phrasing of IKD(Bitcoin).
\end{enumerate}

\subsection{Cryptocurrency (CC)}
Bitcoin is often called a cryptocurrency (CC). The underlying assumption must be that there is a class of so-called cryptocurrencies to which Bitcoin belongs. We criticize the classification on the following grounds: 
\begin{enumerate}
\item A cryptocurrency must be a currency.\footnote{%
This inference is debatable, but we consider the term cryptocurrency confusing it it does not mean something 
roughly equivalent to the following: an
informational currency, complying with generally accepted requirements for 
``currencyness" (or equivalently moneyness), for which control (of agents over 
amounts and corresponding inspections and transactions) has been organized by means of 
encryption and decryption rather than by means of restricted access policies
based on conjectural pseudomonopresence of passwords (see~\cite{BergstraL2013})  or on physical access 
restrictions.} But confirming the status of a ``system" as being a 
currency depends on a plurality of observers some of whom may require that a certain acceptance or usage
must have been arrived at by a system before it can be classified as such. 
Upon its inception Bitcoin did not possess that level of acceptance, and for that reason 
Bitcoin has not started its existence as a cryptocurrency.

\item Being a cryptocurrency is a status that a system may or may not acquire over time.\footnote{%
From~\cite{Barber2012} we quote the question 
``Does Bitcoin have what it takes to become a serious candidate for a long-lived stable currency, 
or is it yet another transient fad?'' This phrasing suggest that Bitcoin might be considered 
a candidate currency, and that a currency need not be either long lived or stable.} 
Assuming that Bitcoin is considered to be a cryptocurrency at some stage then there will most likely be variations (alternative 
designs and systems) of Bitcoin around (perhaps hardly used anymore) which 
have not been that successful. Such alternative systems should 
be given the same type, so that Bitcoin might be considered a successful instance of that type. 
Clearly CC cannot be that type as it contains only systems that have already become successful to a significant extent.

\item Because being a cryptocurrency is the primary success criterion for Bitcoin its 
classification as a cryptocurrency amounts to a value judgement or a quality assessment rather than as 
an initial type.\footnote{%
Classifying every book as a bestseller is a similar mistake, and so is classifying every song 
as a hit.}
\end{enumerate}
We conclude that as an initial top-level classification for Bitcoin CC is not plausible. 
That is consistent with the viewpoint that at some stage (and/or in the eyes of some observers) 
Bitcoin may be or become a system of class CC.

\subsection{Digital currency (DC)}
The ``problem" with digital currency as a member of BT(Bitcoin) 
is similar to the problem mentioned above with CC. 
We prefer IM over DC as in general we prefer money over currency as a 
general term and moreover we consider ``digital"
to carry an implementation bias which ``informational" avoids.

\subsection{Informational money (IM)}
The difficulty with informational money (IM) as an original class for Bitcoin is precisely the same as for
cryptocurrency. Assuming that an informational money is a money, 
the classification expresses a level of achievement that should not be presupposed.\footnote{%
In~\cite{Bergstra2012d} the term informaticology (IY) has been explored. 
In that work Informaticology (IY) is decomposed as: 
IY = CS + DS + FS = Computer Science + Data Science + Fiction Science. IM has roots
in each of these components. The present paper may be classified as a 
contribution to the exploration of informaticology of informational money.}

Our objection to using digital money as an original class for Bitcoin is the same. Stating that Bitcoin can be typed 
as an informational money need not be understood as an acknowledgement of its status as an unlimited success. 
Success is a matter of size of circulation and usage. Success is not meant to imply that no alternative is superior.
\subsection{Informational near-money (INM)}
In~\cite{BergstraL2013}) Bitcoin has been classified as an informational near-money. 
The problem of this classification is that the distance between near-moneys and monies is quite subjective. 
Near-money says no more than ``money-like but perhaps not a money".
Classifying Bitcoin as
an informational near-money rather than as an informational money solves the problem that its 
success in terms of realizing ``moneyness" is not presupposed, but it is unclear to what 
extent ``near-moneyness" is a criterion of partial success of a so-called near money.

\subsection{Virtual money (VM)}
The phrase virtual money shares with informational money and cryptocurrency that it may 
prematurely oversell Bitcoin as a money (or currency). Besides raising that objection to a classification of 
Bitcoin as a virtual money, it is hard to make sense
of virtuality in this setting as like with virtual memory it would suggest that we are not looking at 
real money but merely at a virtual look alike of it. In~\cite{Bergstra2013a} an interpretation of 
virtual money has been given which takes the unreal aspect into account. Nevertheless, that 
particular view of virtual money provides no incentive for classifying Bitcoin as such.

\subsection{Denationalized money (DM)}
A denationalized money is issued by a private party. Although issuing of Bitcoin is very distributed and 
potentially fully anonymous it is not under control of any state. For that reason it merits the qualification denationalized. DM is less plausible as a type because it implies that the status of a money has been achieved.
The criticism regarding DM as a base type for Bitcoin is essentially the same as for CC, DC, and IM.

The Austrian economic school constitutes a source for the viewpoint that monies ought to be denationalized. 
The Austrian school economist Hayek is often seen as an economist the ideas of whom are materializing 
with Bitcoin.\footnote{%
In~\cite{RogojanuB2014}  Bitcoin is examined from the perspective of compliance with principles of the Austrian school.}

\subsection{Bitpenny implementation: with Bitpenny a suitable Abstract Money Type}
In~\cite{BergstraL2013b} we have put forward the suggestion 
that Bitpenny represents an abstract logical specification
of the service that one imagines that Bitcoin ideally will provide. For instance 
Bitpenny achieves by magic that double spending
will not occur, while Bitcoin requires sophisticated machinery to implement that property. Bitpenny would
represent an Abstract Money Type (AMT). 
Abstract Money Types have money structures as implementations in the same
way as Abstract Data Types have data structures as their implementations. Money structures are as 
remote from real money as data structures are from real data. 

Money structures are processes which 
evolve in time. Conceivably a money-structure migrates from the status of a near-money 
to the status of a money and back. Indeed being qualified as a money structure $K_m$ 
compliant with $T_m$ (an AMT, thus $K_m \in |T_m|$) reflects no social acceptance in a particular state 
of the structure (as it operates in the real world at some time $t$) $K^t_m$ of $K_m$ as a money.
$K^t_m$ may be merely classified as being money-like.

In the same way $K_d$ being accepted as a data structure  compliant with ADT $T_d$ 
reflects no social acceptance of a particular state $K^t_d$ of $K_d$ as containing data. $K^t_d$ may be 
merely qualified as a data-like state (of $K_d$).

Apart from the name Bitpenny which is merely a placeholder for better names, a weakness of this typing of
Bitcoin is that Abstract Money Types have yet to be defined and developed. In principle, however, this may be
a way to go about typing systems like Bitcoin on the long run. 

\subsection{Attributes and qualifications}
If a car is of type CAR then a red car may be viewed as an item in the intersection of types 
CAR and RED. However, it 
may be more convincing to consider redness as an attribute (property, or qualification),
which applies to entities contained in types or defined by definitions. Some qualities, or claimed
qualities, of Bitcoin are best viewed as attributes which do not represent types. 

\section{Attributes for or properties of Bitcoin}
We first list some qualifications for Bitcoin which we consider to be less plausible as type descriptions.
When contemplating types for Bitcoin below we do not expect to
see these qualifications mentioned. Such qualifications play a role when moving from a type 
description to narrower type description or to a definition.

\subsection{A system for electronic transactions not relying on trust}
This is an original qualification from~\cite{Nakamoto2008}.

\subsection{P2P system}
Bitcoin was designed as  a P2P system. Currently users of Bitcoin must trust the collective of miners and
the pure P2P status is becoming less obvious.

By stating that Bitcoin is a P2P system not enough information is given about its expected role. 
(We don't speak of objectives as almost 
nothing is known (at this stage) about the objectives of Bitcoin's designers.)

Typing Bitcoin as a open source P2P system provides little progress, and it may even be wrong altogether,
as there seems to be no requirement that Bitcoin clients are realized as open source programs.

\subsection{Reduced product set finance (RPSF)}
In~\cite{BM2011,BM2012} we have put forward the notion of an RPSF in order to 
describe financial systems from
which certain features (and corresponding financial products) are banned.
One might consider Bitcoin an RPSF because it does not allow for governance 
by means of management of the monetary base.

Classifying Bitcoin as an RPSF is not plausible at this stage because like money as well as like  currency,
finance expresses requirements on a system that Bitcoin may not yet meet.

\subsection{The first successful digital coin: a landmark in the history of information security }
One of us has a vested interest in the history of information security (see~\cite{LeeuwB2007}) and
viewing P2P based money-like systems as a historic necessity of which Bitcoin is a first
realistic exemplar may be justified and may on the long run be what will take place. 
What counts against this approach to Bitcoin typing is that it insufficiently expresses 
the important aspect that Bitcoin may still have a long and successful future existence and evolution 
ahead of it (and regarding proposals for its improvement see~\cite{Barber2012}). 
In addition not everybody may consider Bitcoin to be a success as a coin, 
if being a coin counts for anything less than being a currency.\footnote{%
As an open source program Bitcoin is successful already, 
by having many participants tot its P2P system, 
by capturing the imagination of public media, by driving the emergence 
of a dedicated hardware industry for mining, and by forcing all major financial institutions to issue policy
statements about it.}

\subsection{Converting rejected base type options to attributes}
When criticizing IM as a base type for Bitcoin one may uphold that it is an attribute that 
Bitcoin might be a proto IM, where a pro to $X$ is an entity which is expected (hoped, intended) 
to evolve sooner or later into an $X$. Now it is reasonable to have proto CC, proto DC, proto IM, 
and proto DM as attributes of Bitcoin.

\section{Candidate types for BT(Bitcoin)}
Given some kind $X$, which may be an artifact kind, or a natural kind, 
finding a suitable type in BT($X$) requires (i) the preparatory collection of a number of options, 
that is candidate members, (ii) validating the conclusion that BT($X$) is non-empty, and (iii) choosing an optimal
element OBT($X$) of BT($X$).
\subsection{Candidate cryptocurrency (CCC)}
Candidate cryptocurrency (CCC) may well be a type in BT(Bitcoin), and CCC refers to a larger class than CC, while it is contained in MLICcp as defined below. A disadvantage of CCC as a type is that there is no known procedure for leaving the candidate status. Another problem is that some may think that it can be shown that Bitcoin
cannot acquire CC status, from which it would follow that Bitcoin is not a CCC either.
\subsection{Cryptocoinage (CCo) } 
A cryptocoinage (CCo) plausibly refers to the coinage of a larger money-like system. 
As coinage is unlikely to refer to an entire money including each of its money types, 
this is not a plausible typing of Bitcoin. The name Bitcoin is suggestive of it being about coins but we fail to
see in what sense the quantities of Bitcoin are to be viewed as coins.

\subsection{Internet based business (IBB)}
Possession of BTC quantities may be considered as a proportional share in an new kind of  internet startup (now in year 6 of its existence). Mining is the method for issuing and distributing  new shares. New shares are issued at a diminishing rate, and the hypothetical objective of the business is to become dominant factor in what is called today the monetary system. Bitcoin shares constitute the money at the same time. If Bitcoin happens to become a dominant factor in the world-wide monetary system the value of a BTC will rise astronomically (though not expressible in the Euro's from the past anymore).

The business model is to become strong on the internet in an underground style. In~\cite{BergstraL2013} 
it has been argued that this view allows an approach to the determination of the value of a 
BTC. Thinking along this lines the current rate of EUR 600 to a single BTC will correspond with a 20 year 
survival probability of Bitcoin (as an internet based business) of $ 10^{-5}$.

IBB is a reasonable type for Bitcoin if one intends to explain the economic risks, both upside and downside,
of an ``investment" in Bitcoin (i.e. buying BTC against EUR). As a business model IBB is really new, which is an attractive aspect of this candidate type. But we believe that the virtue of IBB as a type for Bitcoin is metaphorical rather than intrinsic.

\subsection{Investment scheme (IS)}
Assuming that the store of value function of Bitcoin outweighs its money of exchange function, one may 
suggest that it constitutes an investment scheme IS). This view is valid if BTC is bought and sold in exchange
for conventional monies only. It seems rather implausible that Bitcoin can survive on this basis, but at the same that very
state of affairs cannot be excluded and for that reason we consider  ``investment scheme"  
to be a conceivable type for Bitcoin.

An individual user viewing Bitcoin as an investment scheme only is currently considered to be a speculator,
because there is no intrinsic argument for BTC to increase its value against conventional monies.\footnote{%
The number of outstanding BTC will have some maximum below 21 million in a phase where mining results
balance losses of Bitcoin accessibility by users who physically use secret keys to active accounts. 
From that point onwards losses (of al users together) will outweigh mining (creation). 
On the very long run, and under the assumption that Bitcoin loss will always have some 
positive probability each BTC will get lost. }

\subsection{Multi-player computer game (MPCG)}
One may consider Bitcoin to constitute a distributed and multi-player computer game (MPCG). 
The game is to acquire control over as many BTC as possible. Good gaming is rewarded by other gamers who
buy accounts. As a side-track all participants collectively try to make progress concerning attacks on the
secure hash function SHA-256.

Viewing Bictoin as an MPCG deviates from a common understanding of its purpose but as a typing 
it may be useful for some purposes.
\subsection{Informational money-like commodity (IMLC)}
Bitcoin provides tradable quantities, with money-like status. 
On that basis it may be viewed as a (system implementing a) money-like commodity (MLC). 
Because its exists (or at least it can can exist)  in terms of information only Bitcoin
is also an informational MLC. We notice that MLC is also a plausible candidate for a type in BT(Bitcoin),
but we consider it to be significantly less informative than IMLC. 
In item~\ref{MLIC} below we will discuss the way in which Bitcoin might be considered to constitute a commodity.

\subsection{Money-like informational commodity (MLIC)}
\label{MLIC}
Viewing Bitcoin as a system providing a platform offering the following features:
\begin{enumerate}
\item a system for giving agents access, and 
\item  facilitating the exchange of that access, to
\item informationally given amounts measured in BTC the unit of Bitcoin), through 
\item the scarce resource of collections of accessible (to the agents) secret keys, and 
\item ``a Bitcoin" as a unit of access within this system, 
\end{enumerate}
one may grasp how Bitcoin may refer to a commodity, 
the substance of which consists of information that is independent of any accidental carrier of it, while 
access to it is scarce.\footnote{%
Informational commodity is an subclass of  information object, a class for which a conceptual 
analysis can be found in~\cite{DoerrT2012}.}

Thus, in more detail, said commodity amenable to being in the possession of an agent, consists of: 
conjectural pseudo-monopresent (that is secret) ``keyware" (that is collections of  keys), 
the size (commodity measure, commodity weight, commodity volume) of which is determined as the 
sum of the amounts these secret keys give access to. 

After an exchange the same amounts are accessible from other secret keys,
equally assumed to enjoy conjectural pseudomono-presence, though in the possession of another agent.

The exchange value of a commodity positively correlates with its size, and with the public trust in the 
pseudo-monopresence (see~\cite{BergstraL2013}) of the (secret) keys from the 
perspective of the agent currently possessing the keyware.

Playing the role of a money, we will speak of a money-like informational commodity (MLIC).\footnote{%
We notice that viewing information as a commodity has a long history, e.g. see~\cite{Detlefsen1984}.
In~\cite{Frow1996} explicit mention is made that information as a commodity will have a price, 
plausibly in excess of the cost of delivery, whereas information in the public domain need not.}

\subsubsection{Ownership and possession of MLIC}
A vast legal and philosophical literature explains concepts like property, title to
ownership, ownership of property, and possession.
We mention~\cite{Epstein1979} as an example of such works. In that paper 
the coming about of property through the rule of first possession is scrutinized. 

We will assume that the commodities underlying an MLIC 
can be in the possession of agents. Possession of a quantity of the commodity 
provides an agent with a number of capabilities that are self-explanatory from the perspective
of a conventional functionality of money.
The extent to which agents can be owners of an item may vary from case to case. By speaking of an
MLIC no commitment is made to a particular bundle of rights which is implied by ownership or possession of
an item. Technically the same MLIC may exist in different legal regimes.\footnote{%
This independence of circulation technology from legal settings is discussed below in 
paragraph~\ref{LOC} on physical coins.}

\subsubsection{Preference of MLIC over IMLC}
As a type for Bitcoin MLIC is  acceptable as well as IMLC, though we prefer MLIC because:
\begin{enumerate}
\item  Bitcoin cannot be an IMLC without being an informational commodity. 
\item The phrase informational commodity is relatively new and without it being 
accepted IMLC is hard to make sense of.
\item Without insisting that Bitcoin keyware is a substance constituting an informational commodity, 
the claim that its secret keys give access to a commodity seems to be unconvincing. 
\end{enumerate}

The use of the concept of a commodity requires justification. 
About the question what is a commodity we take a quote from~\cite{Anderson1990}, which as we claim 
applies to Bitcoin keyware:
\begin{quote} To say that something is a commodity is to claim that the norms of the market are appropriate for
regulating its production, exchange, and enjoyment. To the extent that moral principles or ethical ideals preclude 
the application of market norms to a good we may say that the good is not a proper commodity.
\end{quote}

\subsubsection{Remarks on commodification and informational commodities}
Necessarily an MLIC is a kind of commodity. In other words MLIC is a subtype of commodities.
Commodities may come about in different ways. Individual commodities are build, produced, generated, or 
configured, whereas a specific commodity type may come about through a process of commodification. 
Commodification transforms a type of disparate (though somehow related) entities into a 
class of entities that can be exchanged on a free market and which are paid for by money.\footnote{%
In~\cite{WatsonK1994} commodification is defined as: 
``the process by which objects and activities come to be evaluated 
primarily in terms of their exchange value in the context of trade in addition to any 
use-value such commodities might have."}

We suggest that Bitcoin may be considered a result (among many other such results) of commodification 
(of circulating amounts) starting from digital monies that are still under state control just as conventional
monies like EUR and USD. Such monies seem not to meet the definition of a commodity because free 
market forces are not allowed to regulate their market value.

In~\cite{Gryz2013} the phrase
informational commodity is used for private information which is owned by an individual. The institution of
ownership for that particular kind of commodity is called privacy. As a type of that sort of information we
suggest: personal data oriented informational commodity (PDOIC). 
Privacy indicates that commodities of type BLIC
may be owned by an owner with the effect that all forms of access by non-owners are a breach of the rights
the owner enjoys in virtue of his ownership. The use of informational commodity is justified in this case 
because there is a market for such information, even if privacy is about regulating and constraining that market.

\subsection{Partial informational money (PIM)/partial cryptocurrency (PCC)}
With a partial money we may denote any near-money or money-like commodity which can fulfil some selected subset, 
(possibly all) of the functions required of a money, and which in addition may fulfil these only partially 
what is expected for these selected functions. Given an understanding of an informational money (IM) or a
cryptocurrency (CC), one may derive from this definition of a partial money what a partial IM (PIM) or a partial
CC (PCC) can be. 

Both PIM and PCC are plausible types in BT(Bitcoin) and plausible candidates for the role of an OBT(Bitcoin).

\section{MLIC, our proposed type OBT(Bitcoin)}
The main contribution  of our paper is formulating the proposal that MLIC is in BT(Bitcoin) and 
moreover that MLIC = OBT(Bitcoin).\footnote{%
In \texttt{http://en.wikipedia.org/wiki/List\_of\_cryptocurrencies} for a survey of so-called 
cryptocurrencies, all of which we propose to classify as MLICs also. 
See also~\cite{Steadman2013}.}

\subsection{Motivation}
We see the following arguments in favor of setting OBT(Bitcoin) = MLIC.
\begin{enumerate}
\item MLIC is very compact and it specifies a large class of systems from which Bitcoin might emerge
as the winner.
\item MLIC seems not to make any false or even contested claims.\item MLIC is applicable to Bitcoin during
all stages of its ``life".
\item MLIC is preferred over PIM and PCC because it is ``more consistent" with the eventuality that Bitcoin
may not be classified as a money after all in which case its final type would probably be neither 
PIM nor PCC because arguments for witholding Bitcoin the status of a money would stand in the way. 
\end{enumerate}

\subsection{Weaknesses of the type MLIC in its role of OBT(Bitcoin)}
Typing Bitcoin as an MLIC, however, leads to many subsequent questions some of which may point to
weaknesses of this proposal.
\begin{enumerate}
\item Bitcoin is internet based; MLIC fails to indicate that Bitcoin is only useful for those agents who can access 
the internet freely, to the extent that they can run a specific downloaded client. 
This restriction may be at odds with
property rights on a specific Bitcoin informational commodity to which an agent may be entitled.
\item A potentially problematic aspect of MLIC is the use of commodity as a root  concept. 
Indeed one may claim that
amounts are commodity-like at best but are not commodities proper. This objection may be remedied
by viewing ``money-like commodity" as a semantic unit rather than as  referring to a commodity 
which in addition is money-like. (A similar situation is found with ``free will" which according to 
some philosophers is not pointing to a will (state of willing) that in addition happens to be free.)
\item MLIC fails to express the fact that no backing of value for 
BTC holdings from outside the system is provided.
All value of Bitcoin holdings is to be realized within or by the 
Bitcoin system, through users who have trust and confidence in the system.

\end{enumerate}

\subsection{Counter arguments and risk analysis}
Many arguments against the typing of Bitcoin as an MLIC may be brought forward. 
It seems that legal, economic, monetary, and informaticological perspectives are so 
disparate that it is inconceivable that a single type
proposed for Bitcoin is equally defensible from each of these perspectives.

For instance, while moving from a commodity to a money (and back) may be a matter of evolution from an 
economic perspective, that may not be the case from a legal perspective. 
And once Bitcoin like systems become
mainstream components of computing infrastructures technical names 
(like database, laptop, cloud, computer network) may become more prominent that use-related types.

In~\cite{NimmerK1992} it is argued that information is intangible and that there is a larger 
distance between information 
and its embodiment than with other, more conventional goods. 
This argument hardly applies to a Bitcoin secret key the
possession of which (under the assumption of conjectural pseudo-monopresence) is very much 
linked to its embodiment.
Thus, it must be accepted at least for some informational commodities (e.g. the one's we are 
contemplating in the context of Bitcoin) a notion of possession, and the corresponding
uniqueness of a possessor, make perfect sense.

We consider it implausible that someone would criticize classifying Bitcoin as an MLIC on the
grounds that it is to early too award it that status, for instance in the light of deficient regulation
 (see~\cite{Kaplanov2012}).

Of course some may consider the classification of Bitcoin as an MLIC as being too cautious, 
a possibility 
which brings us to contemplating MLIC maturity levels. Our use of ``informational" may be criticized
for lacking compliance with a preferred philosophy of information (see~\cite{Floridi2011}). 
In~\cite{CourtoisGN2013} a critique of the design of Bitcoin mining is given. Such arguments do not
stand in the way if classifying Bitcoin as an MLIC but may constitute objections against assigning it 
a higher maturity level such as the level IMoE/MfSoV below.

In any case a risk that we see in opting for MLIC in the role of OBT(Bitcoin) is that legal differences
between commodity and money turn out to be rather large. Considered from a legal perspective, 
control 
over a secret key of a Bitcoin address may prove to be more remote from the 
possession of a quantity of some commodity than it proves to be from the possession of an amount of money. 

\subsection{Ramifications of MLIC}
Assuming that MLIC serves as a base type for Bitcoin and Bitcoin like systems one is led to the question
how large this class of systems might be. For instance one may wonder to what extent 
cryptographic techniques are necessary for the realization of each conceivable MLIC. We briefly consider
two conceivable alternative means of implementation.
\subsubsection{Cryptology based MLIC: CBMLIC}
Bitcoin and similar systems are essentially based on an extensive and essential 
use of cryptographic techniques and
access control seems to play a secondary role only. The role of cryptography will stay prominent as 
Bitcoin evolves and the
role of access control may increase. For this reason we consider the type
CBMLIC (for Cryptology Based MLIC)  to be a valid element of BT(Bitcoin). 
CBMLIC is a refinement of MLIC and we don't prefer it over MLIC 
as a candidate OBT(Bitcoin) because it is still possible that each MLIC is cryptology based in which case the addition
of this aspect to a notation for the type would be futile in hindsight.

\subsubsection{Access control based MLIC: ACBMLIC}
It can be imagined that an MLIC makes use of password protected data only. 
That will probably not be an open source P2P system. We suggest that ACBMLIC is coined as a type for which no
instances have been proposed at this stage. If ACBMLIC turns out to be empty then by 
consequence CBMLIC is equal to MLIC.

\section{Alternatives for MLIC}
Although the future of money is probably informational, 
it would be too soon to write off non-informational monies
or non-informational money-like commodities as being,
 no more than a thing from the past. 
When contemplating alternatives to MLIC it becomes important to consider partial monies or rather compartments of monies.
On the physical side, to understand the importance of metallic coins as a component of the Euro system it is 
relatively unimportant that Euro coins have a fixed value in terms of Euro. Small variations can be handled and one may imagine a situation where coins have fluctuating values and a central bank sees to it that these fluctuations are not too big
by means of targeted interventions, which are not meant to stabilize coin value under all circumstances 
and with perfect precision.
\subsection{Physical MLC: PMLC}
The intuition of money is often connected with coins. At a closer inspection coins are fairly difficult to understand. 
In~\cite{Bergstra2013a} an attempt was made to understand ``the logic of coins". 

\subsubsection{Logic of coins}
\label{LOC}
To see the difficulties with understanding a logic of coins one imagines
the task to define forgery. What is a false coin? Is it a coin that fails to comply 
with requirements on its physical constitution, or is it a coin that started its circulation in a fraudulent manner. 
The difference between both approaches is large. Similarly there is a remarkable distinction 
between coins and banknotes resulting from these having unique identification numbers: 
from two identical banknotes at least one must be the product of counterfeit. 
And if no physical anomalies can be established one is forced to accept a 
theory of counterfeit that takes the entire circulation of banknotes into account.

A thought experiment relevant for Bitcoin is to imagine a PMLC which constitutes of a certain range of coins only.
Of particular importance is to grasp the ramifications of ownership, possession, holding, lending, and stealing.
On needs a model where a community of agents makes use of coins. Each coin $c$ can stand in various 
relations to an agent $P$: $P$ may own $c$, $P$ may be in the possession of $c$ (the contrast between ownership
and possession has been analyzed originally in detail by Kant, 
see e.g.~\cite{Williams1977}, and subsequently by many other authors), or $P$ may be a holder of $c$.
At this stage ramifications abound. In~\cite{Merrill1998} four views of property 
(and ownership) are discussed, each of which
may potentially give rise to another view on how coins relate to agents in terms of ownership, possession, and 
capabilities of these. In~\cite{Snare1972} one of these views, so-called multi-variable essentialism, has been
worked out operationally and in detail by giving six rules of conduct that may be considered constitutive 
of the concept of owning a property, while none of these rules is considered absolutely 
necessary in each individual case.  Different bundles of rights may be attached to owners, 
possessors, and holders, of a coin, thus
leading to different legal settings for a coin circulation based on the same circulation mechanics. We expect that
surveying the different options in the case of metallic coins will shed light on the different 
possibilities in this respect for an IM. Investigating this matter seems not to have been carried out in the
existing literature on coin circulation

In~\cite{BergstraL2013} ``exclusively informational money" (EXIM) was ``coined" for denoting an 
IM where ownership is reduced to possession. By looking at metallic coin circulation models, legal 
and philosophical aspects of such ``exotic" legal options (for different relevant bundles of rights) 
can be studied without being detracted by the 
complexities of distributed computing. In~\cite{Holmes1877} intention is put forward as a 
prerequisite for possession, and different views are contrasted
regarding the perspective (intention) of future ownership being a prerequisite for possession as well. Following
this line of thought an EXIM cannot even be defined without making use of a notion of intention, and
in addition  it may hardly have a place in a world dominantly inhabited by artificial agents.

\subsubsection{Advantages of (metallic) coins}
Looking at physical coins as a means of exchange which itself constitutes 
a commodity used exactly for that purpose, the following
advantages of metallic coins are noticeable:
\begin{enumerate}
\item Use for exchange is possible in a significant range of seemingly adverse circumstances: outdoor, windy weather, 
wet and dry, low and high temperature, during a power outage.
\item Use for exchange is technically simple.
\item Safeguarding the physical security of coins is comparable (and no more difficult)  
to maintaining the physical safety of one's body.
\item The possession of coins can be demonstrated to a potential partner before a transaction is performed.
\item Storage is relatively easy and can be performed for a very long time.
\end{enumerate}

We expect that money-like coins will be around for a very long time, in the light of these advantages. 
That does not necessarily imply that such coins must be pegged to entities/quantities in another money-like system
which plays the role of a (real) money. Although these advantages are well-known for metallic coins it is quite possible that
new materials allow for the design and mass production of even better coins.
\subsection{Computational MLC: CMLC}
Computational money unlike informational money has the property that each occurrence of it requires a functioning
machine. For instance even if one has a security code memorized that allows access to some chip card
containing a money like quantity, only the working chip card once activated via the security code will
allow spending the (near) money (or the money-like commodity) that it contains. If that chip card is destroyed the commodity it contains will disappear simultaneously with it.\footnote{%
THe Dutch Chipknip is an example of this.} 

This differs from the situation of Bitcoin where one may imagine that all blockchain data are printed, that all wallet information
is physically stored, that all machines are destroyed and that in a second round every user get a new machine. Subsequently
all data written on paper (or on disk) are loaded into the various machines and the entire Bitcoin system is again live.
With a computational money or more generally with a computational MLIC a corresponding course of events 
is unimaginable.

\section{MLIC maturity levels}
Supposing that an MLIC say XYZcoin progresses from the initial MLIC classification to the stage of  an
informational money (IM). Now having IM status implies that the moneyness of XYZcoin has been 
sufficiently generally accepted, the latter depending on one's philosophy of money.  
It is practical that XYZcoin's progression through the hierarchy of maturity levels towards the IM stage 
moves trough stages that admit explicit naming as well.\footnote{%
In~\cite{Hanley2013} arguments are listed for expecting demise of Bitcoin. Having a maturity level
classification scheme at hand,  such arguments can be framed positively, 
acknowledging what has been achieved, rather than negatively.} In this Section we will propose a 
naming scheme for such maturity levels.

\subsection{Below IM: CIMoE, IMoE}
An informational money of exchange (IMoE) 
may fail as a money of account (MoA) and it may mail fail to 
provide a (reliable) store of value (SoV, or MfSoV, money for SoV), 
it may fail to provide the function of a money of 
documentation (MoD), that is a tool for the preservation of historic information in excess of mere
balance sheet aggregation.\footnote{%
More functionalities of money can be distinguished, for instance MfVM (money for 
vending machines, see~\cite{BamertDEWW2013}.)} 

Before IMoE there is the candidate IMoE stage. An MLIC is a CIMoE
if the path of development towards the IMoE stage is conceptually clear, though there may be 
risks that it will not be successful. We consider Bitcoin to have reached maturity level CIMoE. This
typing judgement implies that we consider it to provide sufficient defense against double spending attacks
(e.g. see~\cite{Herrmann2012}).

\subsubsection{IMoE/MoD, IMoE/MoD/MoA, IMoE/MoD/MoA/MfSoV}
Beyond the IMoE stage we distinguish IMoE/IMoD, where the MLIC serves as an MoE and 
also as an MoD. A further stage is IMoE/MoD/IMoA, which is a stage where an MLIC serves as an 
IMoE/MoD and as an MoA at the same time. Still the long term SoV functionality may be considered
problematic. If that functionality has been acquired as well we find the maturity level
IMoE/MoD/MoA/MfSoV. 

Exactly which features can be productively combined in an informational money requires further
thought. At this stage many issues about the integration of these features and about 
system governance, quality control, and risk management must be considered.\footnote{%
For instance robustness against abuse must be taken into account (see~\cite{Huang2013}).}

\subsubsection{CIM}

An important intermediate stage close to IM is CIM, candidate IM. A CIM is an MLIC that
must have reached IMoE/MoD/MoA/MfSoV maturity. Moreover,  XYZcoin has reached CIM level if
it has become sufficiently clear how XYZcoin can further develop in terms of usage and support structure
so that it would eventually qualify as an IM. Bitcoin has not yet reached CIM status because it is not sufficiently clear that the MfSoV functionality is potentially performed sufficiently 
well by any Bitcoin-like MLIC, including Bitcoin itself. CIMs compete for IM status and achieving IM 
maturity is a gradual matter.

\subsection{IM}
IM status is a very high maturity level. When having reached that status it competes with EUR and USD
on par in terms of technology and mechanisms. 
This stage may be quite distant, it may never arise. And it may be transient as the evolution of an MLIC
may also involve a decrease of maturity level.

For an IM some robustness against risks is important. 
Nevertheless robustness a against risks may be uncertain. The use of  elliptic curve 
cryptography (see ECDSA in \cite{JohnsonMV2001}) makes it vulnerable 
to attacks by means of quantum computers, once in existence. 
We assume that this is not an argument against Bitcoin having a CIM or IM status
at some stage because we assume that as an evolving P2P system it has the flexibility to develop 
towards making use of stronger cryptographic methods when needed. A similar remark would be made
in reply to someone who considers SHA-256 (see~\cite{BergstraM2013} for a recent specification of that algorithm
by means of an instruction sequence) to be a critical weakness of the mining mechanism.

Coins from a conventional money can be handled anonymously. It is unclear to us to what extent the
ability to make anonymous use of a money (or of the transfer of specific items of a particular money) must
be considered ``intrinsic" to a money. And if so, it is unclear to what extent Bitcoin might even achieve 
CIM status, given the deficiencies concerning its anonymity feature which have been widely documented
(e.g. see~\cite{AndroulakiKRSC2012,BritoC2013}).

\subsubsection{What is an IM?}
The quality of being an IM is hard to grasp in a few words. Quite difficult to assess is the quality of an
IM as a store of value. We make some remarks about this particular aspect.
\begin{itemize}
\item The MfSoV (money for store of value) quality
of the Euro depends on one's trust in an open ended political process
currently guided from  institutions seated in
Brussels and in Frankfurt am Main. Suppose one thinks in terms of say 250 years. 
Is the Euro any better than Bitcoin, who knows?
\item Against which disasters, social unrest, earthquakes, inundations, fires, crimes, or other
problems, must a SoV function of money protect the 
holder of an amount? Is transfer to a next generation upon the holder's death an important matter, 
and if so what requirements must be imposed. As it seems this ``problem" is far from having been solved for Bitcoin.
\item One may imagine that metallic coins for Euro's are replaced by money-like commodities, so as to
undo the Euro (or any other recognized money) from the complexity of coin circulation which introduces 
such aspects as loss and wear in ways foreign to the informational world. One may also introduce
physical coins as a money-like commodity to a Bitcoin style MLIC. 
Is this form of convergence between monies a reasonable path to contemplate? 
This matters if one insists that metallic coins (or banknotes) offer their users
a package of advantages that cannot be replaced by means of  information technology based equipment.
\item Is it at all relevant to compare a closed world such as Bitcoin, the evolution of which is
supposedly limited to a fairly restricted scope of technical modification and to minimal external influence
of a political nature, to a money based on an open ended political process (such the Euro). Is it
reasonable to say that the Euro provides an SoV if in fact the EU political process constitutes (or
claims to constitute) that SoV? If the ``value" of money (say in EUR) is merely a claim on the 
outcome of a political process, then store of value is rather empty if the ability of the political system to
materialize the claims is undermined. 
\item If Euro's don't do much more than providing a  bookkeeping of claims which may 
or may not loose their value, that is Euro's are mainly a money of account for measuring the value of bonds, 
then storage of value must take place (if at all) outside the sphere of money.
The objection to an IM (supposedly still seen as a CIM) that it fails to provide a 
store of claims against an important institution (an objection that has been formulated about Bitcoin)  
may be valid while at the same time there is no support for the 
objection that it fails to provide a store of value (which has been formulated about Bitcoin as well). 

For making this distinction between storage of claims and 
storage of values we will put forward two arguments. First of all with a conventional money the institution 
stores the underlying value in different ways (from storing money proper) thus undermining the argument that
a conventional money provides a store of value (and thereby weakening the objection that an IM does not).
Secondly the social acceptance and ubiquitous usage
of an IM may constitute a ``real value" which is stable and may be considered being stored, 
in a way that a conventional money does not.  It is conceivable that precisely because a conventional
money constitutes a bookkeeping of claims its own value (as a communicative tool) is minimal once
the underlying claims are not counted to contribute to the value 

\item In different words an IM (which Bitcoin might become) serves as an important part of
the communication infrastructure of a community and in that form it stores exactly the combined value of that
communication mechanism. Beyond doubt this is a real value and at this stage it is hard to assess 
how it will compare with the value of assets that underly the stability of conventional monies.
\end{itemize}

\subsection{Beyond IM: RDIM, DIM}
Beyond the IM stage we will distinguish two further
development stages of an MLIC: RDIM, a relatively dominant IM, which is an IM that plays a dominant role between 
other IMs (that is relative to the class of MLICs), and a DIM, 
a dominant IM which is an IM that plays a dominant role in comparison with all other monies.

In~\cite{BergstraL2013} EXIM (exclusively informational money) has been put forward as a further important
modality of an IM. Perhaps EXIM may be placed between IM and RDIM, depending on how important
the ``exclusively informational" feature is seen to be.

\section{Concluding remarks}
We have coined MLIC (informational money-like commodity) as a preferred type indication (base type) for Bitcoin. 
We have compared this meta-type to several other candidates for a preferred base type for Bitcoin.

In~\cite{BergstraDV2011} the notion of a conjectural ability has been promoted as a 
means to assess what a proposed theory might achieve in pragmatic terms. 
Viewing our typing of Bitcoin as a theory of Bitcoin, one may
ask which conjectural abilities an appreciation of this theory may create. 
The ability that we assume to be created by 
an appreciation of this (fragment of) theory on Bitcoin is to allow for a systematic discussion 
of its development through all stages including an initial stage and a possible demise without 
being constrained by the implications of it being a money or a near-money.
 
\bibliographystyle{plain}

\end{document}